\documentclass[english]{emulateapj}
\usepackage[T1]{fontenc}
\usepackage[latin9]{inputenc}
\usepackage{graphicx}
\usepackage{amssymb}

\makeatletter

\newcommand{\lyxdot}{.}

\makeatother

\usepackage{babel}

\begin{document}

\title{The old environment of the faint calcium-rich supernova SN 2005cz}

\author{Hagai B. Perets$^{1}$, Avishay Gal-yam$^{2}$, R. Mark Crockett$^{3}$,
Joseph P. Anderson$^{4}$, Phil A. James$^{5}$, Mark Sullivan$^{3}$,
James D. Neill$^{6}$ and Douglas C. Leonard$^{7}$}

\affil{$^{1}$Harvard-Smithsonian Center for Astrophysics, 60 Garden St.,
Cambridge MA 02338, USA\\
 $^{2}$Weizmann Institute of Science, POB 26, Rehovot, Israel\\
 $^{3}$Department of Physics, University of Oxford, Denys Wilkinson
Building, Keble Road, Oxford OX1 3RH, UK\\
 $^{4}$Departamento de Astronomia, Universidad de Chile, Casilla
36-D, Santiago, Chile\\
 $^{5}$Astrophysics Research Institute, Liverpool John Moores
University, Twelve Quays House, Egerton Wharf, Birkenhead CH41 1LD,UK\\
 $^{6}$California Institute of Technology, 1200 E. California
Blvd., Pasadena, CA 91125, USA\\
 $^{7}$Department of Astronomy, San Diego State University, San
Diego, CA 92182, USA}
\begin{abstract}
The supernova SN 2005cz has recently attracted some attention, due
to the fact that it was spectroscopically similar to type Ib supernovae
(SNe), a class that is presumed to result from core-collapse of massive
stars, yet it occurred in an elliptical galaxy, where one expects
very few massive stars to exist. Two explanations for this remarkable
event were put forward. Perets et al. (2010) associate SN 2005cz with the
class of Ca-rich, faint SNe Ib, which likely result from old double-white-dwarf
systems with a He-rich secondary. On the other hand, \cite{kaw+10}
suggest that SN 2005cz is indeed a core-collapse event (in a binary system), 
albeit of a star at the lower end of the mass range, $10-12\,M_{\odot}$. 
The existence of this star in its elliptical
host is explained as resulting from low-level star formation (SF) activity
in that galaxy. Here we present extensive observations of the location
of SN 2005cz, sensitive to a variety of SF tracers, including
optical spectroscopy, $H\alpha$ emission, UV emission and HST photometry.
We show that NGC 4589, the host galaxy of SN 2005cz, does not show any signatures
 of a young stellar population or recent SF activity either close to or far from the location of SN 2005cz.
\end{abstract}

\section{Introduction}

Recently a novel type of faint supernovae (SNe) with peculiar properties
has been discovered. A group of eight such events has been identified,
all spectroscopically similar to type Ib SNe, but faint (typical
absolute peak magnitude of $\sim-15$) and calcium rich \citep{per+10}.
Although type Ib SNe are generally thought to result from core-collapse
of massive stars (e.g. \citealt{fil97}), a large fraction of the host
galaxies of these faint, Ca-rich SNe are early type galaxies. Additionally,
the ejecta mass of SNe in this subclass appear to be very low (e.g.
$\sim0.3\,M_{\odot}$ found for SN 2005E, \citealp{per+10}; and
$\lesssim1\,M_{\odot}$ found for SN 2005cz, \citealp{kaw+10}),
less than expected and observed for core-collapse SNe of any type.
These SNe were therefore suggested to originate from a different process
involving the thermonuclear explosion of a helium-shell on a white
dwarf \citep{per+10,she+10,wal+10}. Nevertheless, an alternative
scenario involving a core-collapse of a  $10-12\,M_{\odot}$ star,
 which is a part of a binary, was suggested by 
\cite{kaw+10} for the origin of one of the Ca-rich SNe; SN 2005cz.
Here we study this possibility, and look for any evidence for SF 
or young stellar population near the location of SN 2005cz. In the following 
we discuss our results from observations of various SF tracers 
including $H\alpha$ emission, HST photometry, host galaxy spectra and UV emission. 
In addition we shortly discuss the star formation and merger history of the host 
galaxy.  
\section{Star formation traces}

Massive stars are usually formed and observed in giant molecular clouds
and young stellar clusters or associations \citep{chu+08,sch+08}.
Core-collapse SNe from massive stars are therefore expected to be
found close to star-forming regions (SFRs), and overall in SF galaxies. 
We therefore searched for 
SF tracers in the host galaxy both near and far from the location of SN 2005cz.

SFRs produce two classes of emission: continuum emission
from young stars and emission lines (dominated by H$\alpha$) produced
by ionized gas. We have searched for both classes of emission, and
obtained upper limits on the SF rates. In addition we
used HST images to look for nearby point sources which may correspond
to young stars close to the SN location.

\subsection{R-band and H$\alpha$ imaging}
 SN 2005cz was discovered on July 17th 2005, in the elliptical galaxy
NGC 4589 \citep[RA 12:37:27.85; dec 74:11:24.5]{dim+05}.
Published H$\alpha$ narrow-band imaging data exists for this galaxy 
\citep{gou+94}. These data do not suggest any SF activity
close to the location of SN 2005cz, although \citeauthor{gou+94}
do find a significant H$\alpha$+{[}NII{]} emission from the central
regions of the galaxy, aligned along the minor axis where a dust lane
is observed (see Fig. 1 and their Fig. 22).

We obtained an additional H$\alpha$ image of NGC 4589 with the Liverpool
Telescope (LT), and then analyzed it using similar methods to those
described in detail elsewhere \citep{and+09}. We detected 
the dust lane mentioned
above as a perturbation to the isophotes in
our LT R-band image. Notice that according
to \citet{gou+94} the H$\alpha$+{[}NII{]} emission is associated
with the LINER nucleus, and SF is not discussed in that
context; this is further supported by \citet{ho+97a} and our analysis
of the galaxy spectrum (see below and in Fig. \ref{fig:The-nucleus-subtracted}).
Moreover, the emission runs from the galactic nucleus along an axis
offset to that containing the SN 2005cz position. Therefore, there
is no evidence in either the \citet{gou+94} data or in ours for any
line emission near the SN location and even to large distances from it 
(> 1.5 kpc, see also Fig. \ref{fig:The-nucleus-subtracted}). 

Although core-collapse SNe do not locally trace the H$\alpha$ emission from
SF regions, they do trace larger scale SF H$\alpha$ emission in the host galaxies
 of core-collapse SNe \citep{and+09}. In particular, none of the $>100$ SNe in the 
\cite{and+09} sample were found in elliptical galaxies, and only $<3\,\%$ 
of all CC SNe in their survey show H$\alpha$ emission at similar or larger distances 
from the SNe location as found for SN 2005cz.  We note that in the latter cases 
the emission was clearly related to SF, where as the H$\alpha$ emission we find 
in the host galaxy of SN 2005cz (but far from it) is likely related to the LINER nucleus, rather than 
to any SF activity (see next section).  

We also used our data to determine a $3\sigma$ upper limit for the
H$\alpha$ flux from the variation of the sky background, using a
$2''$ aperture, of $1.3\times10^{-14}$ erg cm$^{-2}$ (determined
using a host galaxy R-band magnitude taken from \citealp{san+78}).
At the distance of NGC 4589 ($29.42$ Mpc; taken from NED%
\footnote{http://nedwww.ipac.caltech.edu/%
}) this corresponds to an H$\alpha$ luminosity of $1.33\times10^{39}$
erg s$^{-1}$. Correcting this for Galactic extinction (0.075) and
the contribution from NII lines we get an H$\alpha$ flux of 
$1.07\times10^{39}$ erg s$^{-1}$ cm$^{-2}$ .
Using the calibration from \citet[Eq. 2]{ken98} we find an upper
limit on the SF rate of $SFR_{limit}=8.45\times10^{-3}\,M_{\odot}$
yr$^{-1}$.

\begin{figure}
\includegraphics[scale=0.38]{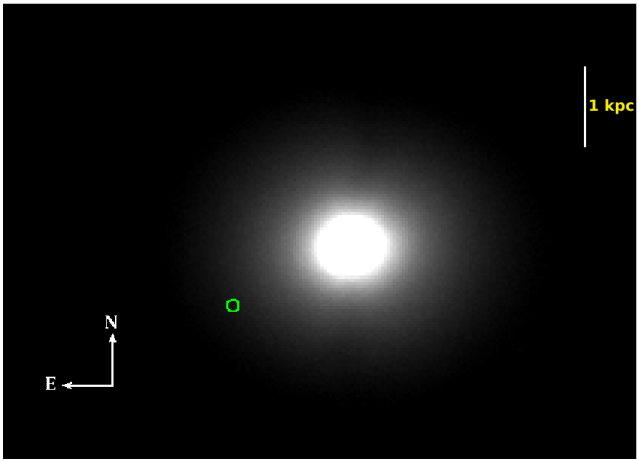}

\caption{\label{fig:R-band}R-band imaging of SN 2005cz host galaxy NGC 4589.
The dust lane reported by \citet{gou+94} is weakly seen here as well 
far away from the reported location of the SN (marked with a circle).}

\end{figure}

\subsection{Host galaxy spectrum}

\begin{figure}
\includegraphics[scale=0.35]{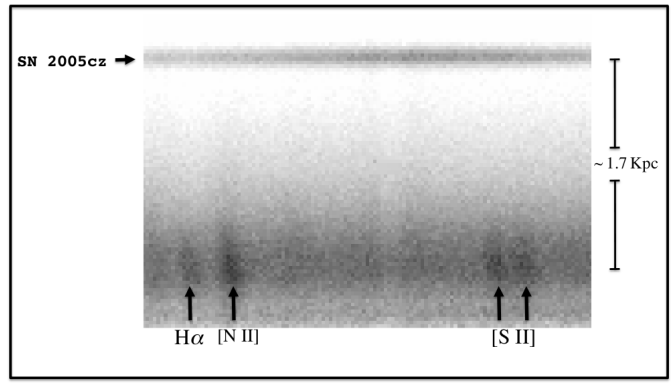}

\caption{\label{fig:The-nucleus-subtracted}
Close-up of the two-dimensional spectrum of SN 2005cz, obtained in a 500
     second exposure at the 10-m Keck I telescope on 28 July, 2005.  The slit
     was oriented at a position angle of $100^\circ$ east of north, centered on
     SN 2005cz. Only the lower portion of the spectral image, which included
     the portion near to the galaxy's nucleus, is displayed here; the upper
     portion showed no emission-line activity.  Arrows indicate the locations
     on the CCD chip of SN 2005cz (horizontal line), as well as the H$\alpha$,
     [N II] $\lambda 6583$, and [S II] $\lambda\lambda 6722/6734$ emission
     lines from the galaxy nucleus.  The relative strengths of the emission
     lines are consistent with the LINER classification given by  \citet{ho+97a}    
      for the nucleus of NGC~4589, and the lack of perceptible
     H$\alpha$ emission close to the location of SN 2005cz in this deep
     spectral image is consistent with the absence of any other indicator of
     ongoing SF activity in the galaxy.}
\end{figure}

We observed the location of SN 2005cz on 28 July, 2005 
using the LRIS spectrograph mounted on the Keck 10m telescope on 28 July, 2005, 
and obtained data from a 500 s exposure. 
The deep, 2-d spectral image shows no  emission features within at 
least 1 kpc of the location of SN 2005cz (see Fig. \ref{fig:The-nucleus-subtracted}).  
While emission lines from the bright galaxy nucleus are seen on
the spectral image, they are $\sim 12.1\arcsec$ ($\sim 1.7$ kpc) away, and
are not those of typical HII regions (e.g., they include strong [S II]
6722/6734); rather, they present line ratios characteristic of a LINER
galaxy nucleus, consistent with the LINER classification of this galaxy
given by \cite{ho+97a}.  Thus, while H$\alpha$ emission is observed in
the galaxy, it is not near the SN location, and in any event shows no
obvious indication of being associated with SF activity.

\subsection{HST photometry }

Post-explosion HST data exist for the location of SN 2005cz (taken
on Nov 11th 2006 with the ACS/WFC F435W, F555W and F814W filters and
exposure times of 1500, 1500 and 1600 s, respectively). These observations
were made as part of HST program GO 10498 (PI: S. Smartt). Photometric
analysis of the objects close to the location of SN 2005cz has been
performed using PSF fitting procedures within IRAF DAOPHOT.

\begin{figure}
\includegraphics[scale=0.65]{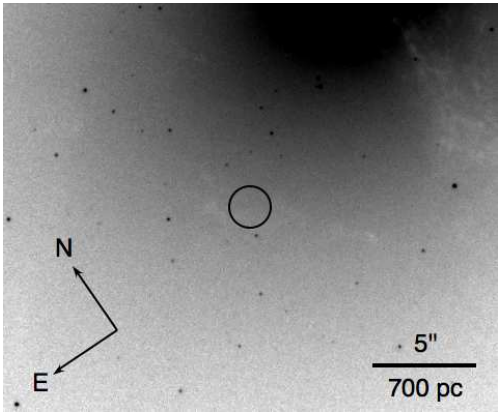}

\caption{\label{fig:HST-photometric-imaging}Post-explosion HST imaging of
the location of SN 2005cz. The image (taken on 2006 Nov 11 - ACS/WFC
F435W - $1500$ s exposure) shows point sources closest to the circled SN
location. The circle has $1''$ radius ($\sim140$ pc) and
is centered on the reported position of SN 2005cz. No sources are
observed within this region. The observed point sources further away
are consistent with being globular clusters (see text).}

\end{figure}

Fig. \ref{fig:HST-photometric-imaging} shows a section of the post-explosion
ACS F435W image. The plotted circle has $1''$ radius ($\sim140$ pc)
and is centered on the reported position of SN 2005cz. As can be seen
we found no object suggesting the existence of massive stars nearby
the reported location of SN 2005cz. The closest source we observe
is  $\sim1.5''$ ($\sim210$ pc, projected; see Fig. \ref{fig:HST-photometric-imaging});
the V-I colors ($\sim1.0$) and absolute magnitudes (ranging between $-7$ and $-9$ mag) 
of objects within $6''$ ($\sim850$ pc) are all consistent with those of globular clusters, 
as seen in other nearby elliptical galaxies (e.g. M87; \citealp{kun+99}).

One of the scenarios discussed by \citet{kaw+10} suggested the progenitor of SN 2005cz 
had been a massive star ($10-12\,M_{\odot}$) with a binary companion. Massive stars
usually have similarly high mass companions \citep[e.g.][]{kob+07}. We therefore
initially attempted to find a detection of the (likely) massive companion
of the SN progenitor in the post-explosion data, but no point source
was visible at the SN site. We therefore derived detection limits
for each image, in an attempt to constrain the luminosity, and hence
mass, of the unseen companion of the progenitor star, or alternatively,
of any other massive star likely to exist in this region if it were
a SFR environment.  Photon counts at the $5\sigma$
level were defined for an aperture of $4$ (ACS/WFC) pixels radius.  Aperture 
corrections (from a 4 pixel to infinite aperture) were taken from \citet{sir+05}, 
while Vegamag zeropoints were taken from the ACS website hosted by
STScI.%
\footnote{http://www.stsci.edu/hst/acs/analysis/zeropoints%
}  The apparent magnitude limits were converted to absolute
magnitude limits using the average distance modulus from NED: $m-M=32.27\pm0.62$
mag and the foreground extinction from \citet{sch+98}: $E(B-V)=0.028$
assuming the reddening law of \citet{car+89}. We then find $5\sigma$
detection limits of $M(F435W)=-6.24\pm0.62;\,M(F555W)=-6.85\pm0.62;$
and $M(F814W)=-7.95\pm0.62,$ where the errors are dominated
by the uncertainty in the distance modulus.

\begin{figure}
\includegraphics[scale=0.6]{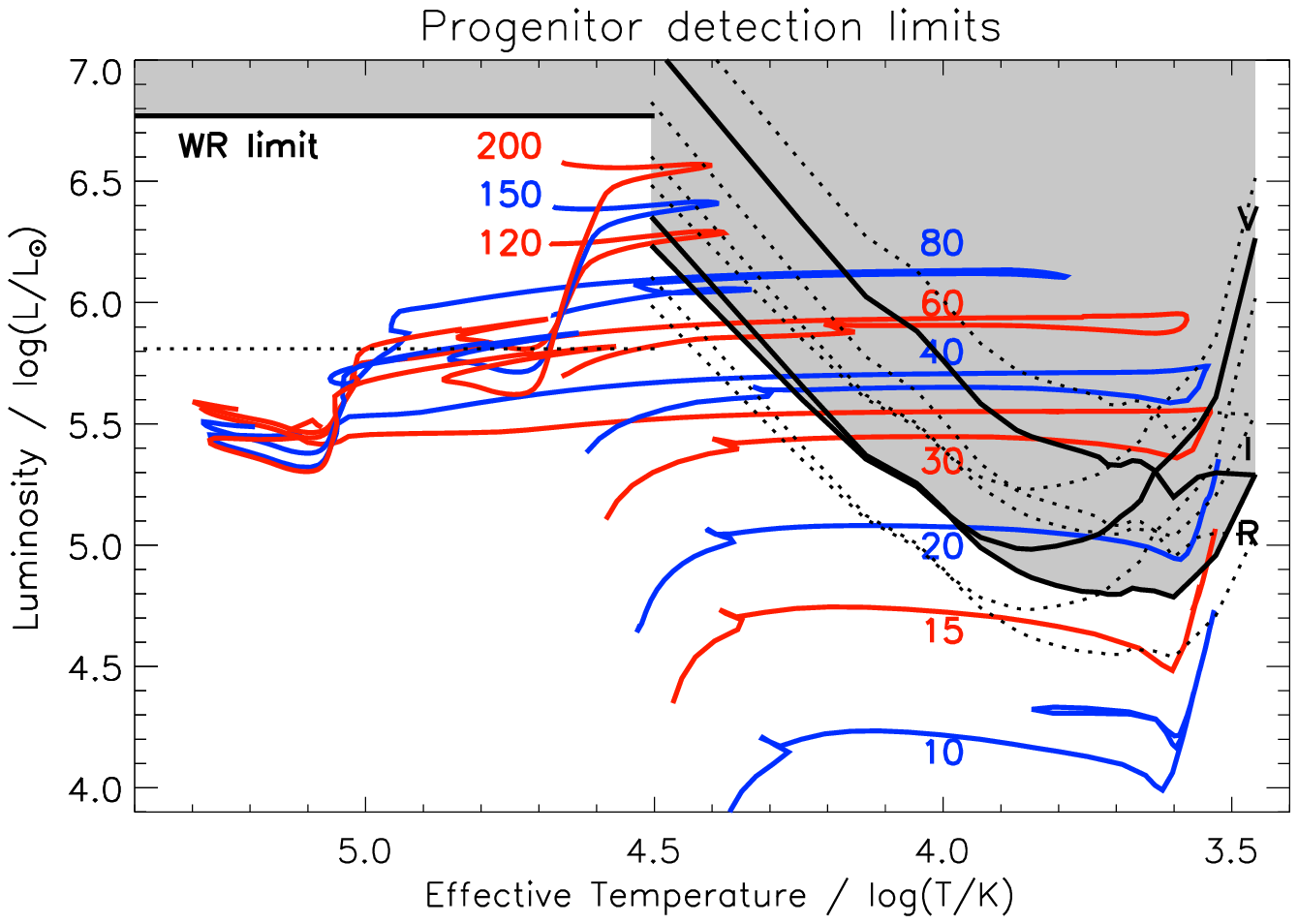}
\includegraphics[scale=0.6]{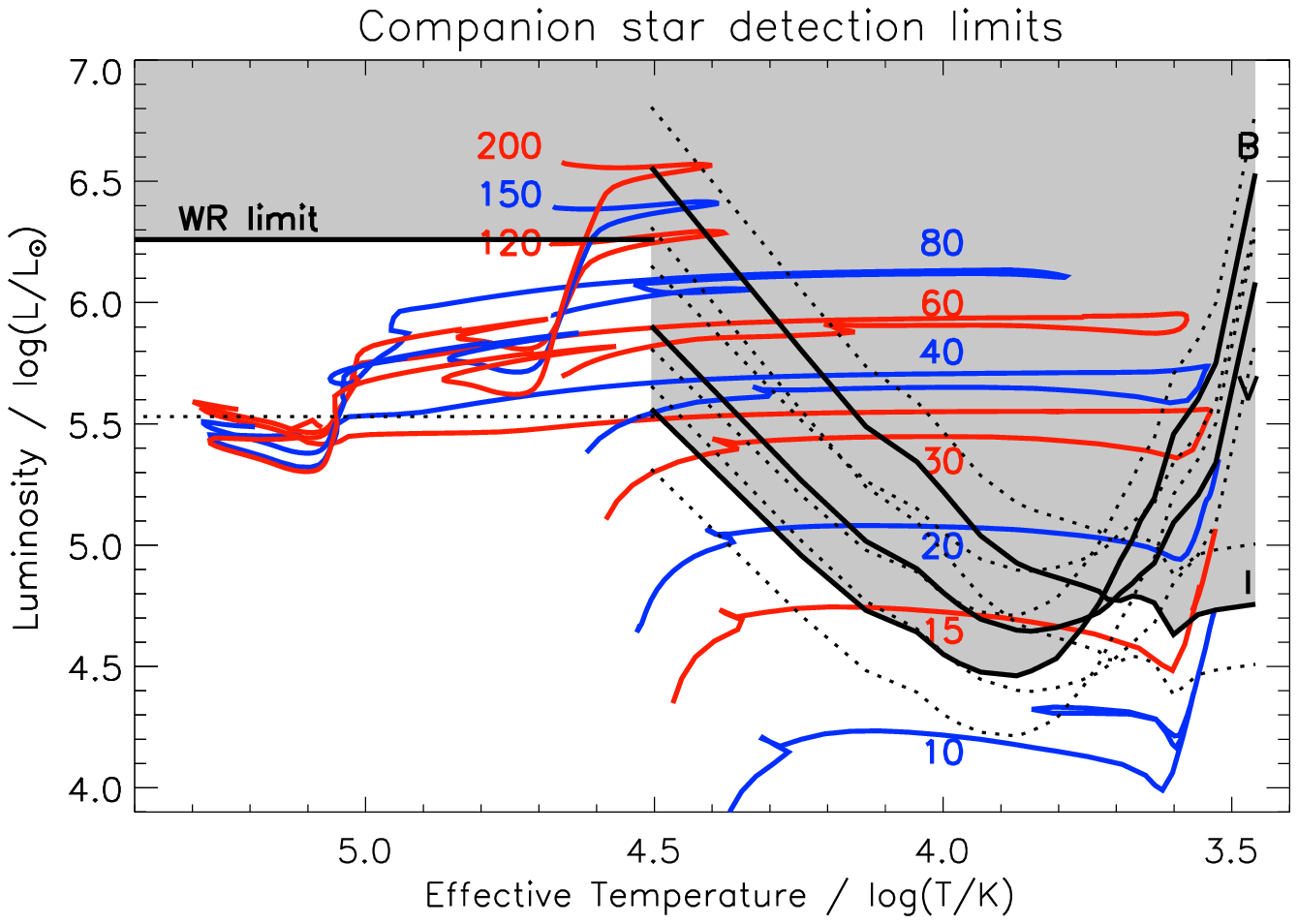}

\caption{\label{fig:HRD} H-R diagrams detailing the spectral types and masses of stars that are excluded (shaded region) by the detection limits
from the pre-explosion (top) and post-explosion (bottom) HST data.
See text for detailed explanations. Also plotted are stellar evolutionary
models created using the Cambridge STARS code \citep{eld+04}. }

\end{figure}

We similarly use pre-explosion HST data from the WFPC2 instrument
which exist for the location of SN 2005cz. In principle, the pre-explosion
data can be used to constrain not only the type of the companion of
the progenitor star (or other massive stars in its environment), but
also the SN progenitor itself. Unfortunately, the pre-explosion data
are of significantly lower depth than post-explosion observations.
These data were taken on May 14th 1994 (GO 5454 - PI: M. Franx), with
the F555W and F814W filters and exposure times of 1000 and 460 s,
respectively, and on January 10th 1999 (SNAP 6357 - PI: W. Jaffe), 
with the F702W filter and an exposure time of 1000 s.  We derived detection 
limits for these images using the same procedure as above;  $5\sigma$ photon 
counts defined for a 4 (PC) pixel radius aperture, with 
aperture corrections from \citet{hol+95}, and the updated 
Vegamag zeropoints of \citet{dol00} taken from his website.%
\footnote{http://purcell.as.arizona.edu/wfpc2\_calib%
} In this case we find detection limits of 
$M(F555W)=-7.70\pm0.62;\,M(F702W)=-7.77\pm0.62;$
and $M(F814W)=-9.32\pm0.62,$ where again the the errors are
dominated by the uncertainty in the distance modulus.

The absolute magnitude limits we find are converted to luminosity
limits for a range of Wolf-Rayet and supergiant spectral-types. To
do so requires a color and bolometric correction for each spectral
type in each of the HST instrument and filter combinations. This is exactly the same approach
used by, for example, \citet{mau+05} and \citet{cro+07}. Here, we
have used the color and bolometric corrections for O9 to M5 supergiant
stars from \citet{dri+00}. The color transformations between the
ACS Vegamag to Johnson-Cousins photometry was carried out using the
methodology and coefficients from \citeauthor{sir+05} (2005) for
the post-explosion data. For the pre-explosion data correction from
the HST flight system to Johnson-Cousins magnitudes were done following
\citet{dol00} and using the updated coefficients from his website$^{2}$.
The results are shown in Fig. \ref{fig:HRD}. The solid, black curves
to the right of the H-R diagrams (HRD), which are labeled B, V, R
and I denote the luminosity limits derived from the F435W, F555W,
F702W and F814W images, respectively. The dashed curves are the $1\,\sigma$
uncertainties of these limits.

For Wolf-Rayet stars we have taken a simplified approach, 
following \cite{cro+07}.  In this case, the F555W and F435W filters
provide the most restrictive limits for such hot stars in the pre-
and post- explosion data (respectively), and these are the only limits
that are plotted on the HRDs, shown as horizontal black lines in the top-left.
 These lines denote the WR luminosity limits calculated
from the F555W and F435W absolute magnitude limits, and using a bolometric
correction of $-4.5$ \citep{smi+89}. The lower, horizontal dashed
lines mark the luminosity limits assuming a lower limit to the bolometric
correction of $-2.7$ \citep{cro07} and the lower bound of the distance
modulus.

Also plotted on the HRDs are stellar evolutionary models created using
the Cambridge STARS code \citep{eld+04}. These models are of solar
metallicity ($z=0.02$) and range in mass from $10\,M_{\odot}$ to
$200\,M_{\odot}$. WR mass-loss rates including scaling with
metallicity are from \citet{eld+06}.

The HST images show no evidence of red supergiants more massive
 than $\sim15\,M_{\odot}$ in the vicinity of the SN location. 
If the progenitor of SN 2005cz
was a $10-12\,M_{\odot}$ star (lifetimes of $18-26$ Myr; \citeauthor{sch+92} 1992), 
we might still have been able to find more massive stars (e.g. 15 $M_{\odot}$, with lifetime of $\sim13$ Myr), given a few Myr age spread inferred for SF regions \cite[e.g.][and references therein]{dar+10,jef09}.  

We conclude that the HST data constrain the possibility of 
recent SF activity, and disfavor a very massive stellar progenitor for SN 2005cz 
(i.e. likely excluding the possibility of a single massive progenitor, 
as suggested in some models for type Ib SN). They also exclude the existence 
of a massive ($\simeq15\,M_{\odot}$) red supergiant binary companion 
to the progenitor of SN 2005cz.  However, these data can not, 
by themselves, strongly constrain a core-collapse model with a $10-12\,M_{\odot}$ progenitor 
for the SN. 
  
\subsection{GALEX, 2MASS and RCS photometry}

\subsubsection{Star formation history of the host galaxy}

\begin{figure}
\includegraphics[scale=0.4]{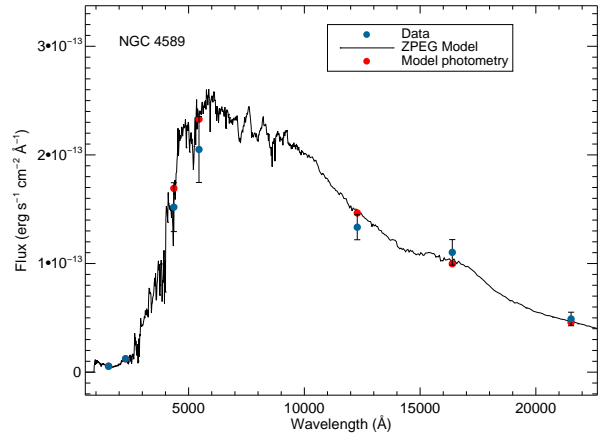}\caption{\label{fig:SFR-fit}A spectrum of NGC 4589, fitted using the Z-PEG
code. The best fit parameters are a total stellar mass of $10^{11.17}\,M_{\odot}$,
\emph{null} specific SF rate, and mean age of $12.5$
Gyr.}

\end{figure}

We use host galaxy photometry compiled in a similar way to that described
by \citet{nei+09}, including ultraviolet data from the GALEX (Galaxy
Evolution Explorer) satellite, and optical photometry from the third
reference catalog of bright galaxies (RC3; \citealp{cor+94}) and
2MASS \citep{skr+06}. We analyzed the data in order to estimate the
SF history and age of NGC 4589 using the methods described
in detail by \citet{sul+06,sul+10}. This makes use of the photometric
redshift code Z-PEG \citep{leb+02} based upon the PEGASE.2 spectral
synthesis code \citep[e.g.][]{fio+97}. Our best fitted spectrum is
shown in Fig. \ref{fig:SFR-fit}.

We find that the photometry of NGC 4589 is best fitted with a total
stellar mass of $10^{11.17}\,M_{\odot}$ (acceptable solutions are
in the range $10^{11.15}-10^{11.19}\,M_{\odot}$), \emph{null} specific
SF rate, and a mean age of $12.5$ Gyr\footnote{Note that the given error ranges are only statistical, and the systematics could be
larger.},  i.e. we find no
trace of any recent SF activity in this galaxy, and generally
infer from the data an old age for the stellar population of NGC
4589.

\subsubsection{Limits on local star formation from GALEX UV photometry}

We have estimated a limit on the SF rate near the SN location
using the measured NUV luminosity. We have relied on the relation
from \citet{ken98} converted to the \citet{kro01} IMF:\begin{equation}
SFR=1.0\times10^{-28}L_{\nu,\, NUV},\label{eq:kennicutt}\end{equation}
 where $L_{\nu\, NUV}$ is the NUV luminosity in units of ergs s$^{-1}$
Hz$^{-1}$, and SFR is the SF rate in $M_{\odot}$ yr$^{-1}$.

We note that SFRs derived from the NUV luminosity for red sequence
galaxies are problematic (see discussion in \citealp{wyd+07}), because
the NUV band can include light from older stellar populations \citep{yi+05,ric+05}.
In such cases, the NUV luminosity will overestimate the recent SFR.
The SFR we find therefore serves only as an upper limit. For an aperture
of $0.5$ kpc around the SN location we find the Galactic extinction
corrected AB NUV magnitude of $22.59\pm0.06$ mag. At the distance of
NGC 4589 this translates to $SFR_{limit}=4.3\times10^{-4}\,M_{\odot}$
yr$^{-1}$. At such low SFR, massive stars are unlikely to form. 
Indeed, as discussed by \cite{pfl+07}, the upper limit for the 
stellar clusters mass function decreases with lower SFR. 
The fraction of ionizing
massive stars is higher in heavy star clusters than in light star
clusters, due to the relation between the mass of a star cluster and
its most massive star.  Therefore \cite[][see their figure 4]{pfl+07} 
find that at SFRs as low as we
find, massive stars of $\sim10\,M_{\odot}$ as required for a core-collapse
SN, rarely form, if they form at all. 

\subsection{Galaxy morphology}
 In support of one of the model in which SN 2005cz originated from a core collapse of 
a young massive star ($10-12\,M_{\odot}$), \cite{kaw+10} suggest that NGC
4589 could have suffered a recent merger, triggering some SF activity. 
However, the morphology of NGC 4589 is relaxed \citep{sch+90}. Even the
dust lane mentioned above is observed at a very low level, and therefore
a major merger occurring 100 Myr ago or less is unlikely.
Although NGC 4589 does have a complex stellar rotation field, its morphology
shows a smooth optical profile following 
the de Vaucouleurs law \citep{moe+89}; 
this could support the possibility that the galaxy is a relatively old
merger remnant; however, this is irrelevant for the discussion of
recently triggered SF. Moreover, even if a minor merger
event occurred in the more recent past, the evidence from the dust lane (Fig.
\ref{fig:R-band}) would suggest that it occurred along the minor
axis, far from the SN location.

\section{Discussion and Summary}
In this letter we studied the local and global environment of SN 2005cz in the elliptical galaxy NGC 4589. We used various SF tracers including optical 
spectroscopy, $H\alpha$ emission, UV emission and HST photometry. 
We also reviewed the the merger history of the host galaxy. 
We found that although some H$\alpha$ emission 
(which in principle could trace SF activity) exists in the host galaxy, 
it is far (>1.5 kpc away) from, and unrelated to the close environment of SN 2005cz. Moreover,
this emission is more likely to be associated with an AGN 
in the nucleus, rather than trace SF activity. Other star
formation tracers (HST imaging of young massive stars, UV, R-band imaging and host galaxy spectrum) 
show no evidence for SF in the galaxy, and particularly
close to the reported location of the SN. The UV emission data could trace 
 stars down to lower mass than H$\alpha$ emission \cite[e.g.][]{gog+09}. 
Therefore, while H$\alpha$ emission may not be detected in older SF regions, 
in which core-collapse SNe from progenitors of $10-12\,M_{\odot}$ 
may explode, they should still present significant UV emission.
The lack of such UV emission therefore suggests that recent SF activity 
has not occurred in this galaxy, and in particular close to the 
location of SN 2005cz.  In addition, the overall structure and 
colors of the host galaxy show no evidence of recent SF
in the last Gyr. The HST data exclude very 
massive progenitors, suggested to be the progenitors of type Ib SNe, and
show no evidence for massive young clusters or $>15\,M_{\odot}$ 
supergiants near the SN location.  

In principle, a massive progenitor could have formed
far from the observed SN location, and later 
have been ejected at high velocity
to explode far from its birth place. Since we find no evidence for
SF even up to 1 kpc from the SN location, such a star should
have been a runaway star to form so far (velocities
of 30-100 km s$^{-1}$, for a lifetime of 10-40 Myr). According to
\cite{kaw+10} the progenitor is suggested to be a binary star. However,  runaway
(or hypervelocity) binary stars, especially massive ones as required
for a SN progenitor, are rare \citep{leo+90,per09b,per+10,per+10b}.
Therefore, although possible in principle, such a scenario would require
fine tuned conditions. 

Taken together, the analysis of UV and H$\alpha$ emission, 
the spectrum of the host galaxy
NGC 4589,  the 2MASS and RCS photometry as well as out HST data, 
show no evidence for recent SF 
near the location of SN 2005cz or even at large distances from it.

We conclude that our results strongly disfavor a young massive-star progenitor 
for SN 2005cz. These results are consistent with the host galaxy type
of other Ca-rich faint type Ib SNe, found to be biased towards early
type galaxies \citep{per+10}. Moreover, the only SNe of types II or Ib/c 
ever to be found in elliptical galaxies (SNe 2000ds and 2005cz) are both faint Ca-rich type Ib SNe similar to SN 2005E \citep{hak+08,per+10}. 
This provides additional support to the suggested origin
of these SNe from a helium detonation in a WD-WD binary system, i.e. 
from a low mass old progenitor rather than a core-collapse of a young
massive star.

\acknowledgments{We gratefully acknowledge NASA, the Centre National d'Etudes Spatiales of France and the Korean Ministry of Science and Technology 
for the development, construction, operation, and science analysis of the GALEX (Galaxy Evolution Explorer) mission.}

\bibliographystyle{apj}

\begin{thebibliography}{44}
\expandafter\ifx\csname natexlab\endcsname\relax\def\natexlab#1{#1}\fi

\bibitem[Anderson \& James(2009)]{and+09} 
{Anderson}, J.~P. \& {James}, P.~A. 2009, \mnras, 399, 559 

\bibitem[{{Cardelli} {et~al.}(1989){Cardelli}, {Clayton}, \& {Mathis}}]{car+89}
{Cardelli}, J.~A., {Clayton}, G.~C., \& {Mathis}, J.~S. 1989, \apj, 345, 245

\bibitem[{{Chu} \& {Gruendl}(2008)}]{chu+08}
{Chu}, Y.-H. \& {Gruendl}, R.~A. 2008, in Astronomical Society of the Pacific
  Conference Series, Vol. 387, Massive star formation: Observations confront
  theory, ed. H.~{Beuther}, H.~{Linz}, \& T.~{Henning} (ASP), 415

\bibitem[{{Corwin} {et~al.}(1994){Corwin}, {Buta}, \& {de
  Vaucouleurs}}]{cor+94}
{Corwin}, Jr., H.~G., {Buta}, R.~J., \& {de Vaucouleurs}, G. 1994, \aj, 108,
  2128

\bibitem[{{Crockett} {et~al.}(2007)}]{cro+07}
{Crockett}, R.~M. {et~al.} 2007, \mnras, 381, 835

\bibitem[{{Crowther}(2007)}]{cro07}
{Crowther}, P.~A. 2007, \araa, 45, 177

\bibitem[Da Rio et al.(2010)]{dar+10} 
Da Rio, N., Gouliermis, D.~A., \& Gennaro, M.\ 2010, \apj, 723, 166 

\bibitem[{{Dimai} {et~al.}(2005)}]{dim+05}
{Dimai}, A. {et~al.} 2005, \iaucirc, 8569, 1

\bibitem[{{Dolphin}(2000)}]{dol00}
{Dolphin}, A.~E. 2000, \pasp, 112, 1397

\bibitem[{{Drilling} \& {Landolt}(2000)}]{dri+00}
{Drilling}, J.~S. \& {Landolt}, A.~U. 2000, {Normal Stars}, 381--+

\bibitem[{{Eldridge} \& {Tout}(2004)}]{eld+04}
{Eldridge}, J.~J. \& {Tout}, C.~A. 2004, \mnras, 348, 201

\bibitem[{{Eldridge} \& {Vink}(2006)}]{eld+06}
{Eldridge}, J.~J. \& {Vink}, J.~S. 2006, \aap, 452, 295

\bibitem[Jeffries(2009)]{jef09} 
Jeffries, R.~D.\ 2009, IAU Symposium, 258, 95 

\bibitem[{{Filippenko}(1997)}]{fil97}
{Filippenko}, A.~V. 1997, ARAA, 35, 309

\bibitem[{{Fioc} \& {Rocca-Volmerange}(1997)}]{fio+97}
{Fioc}, M. \& {Rocca-Volmerange}, B. 1997, \aap, 326, 950

\bibitem[Gogarten et al.(2009)]{gog+09} 
Gogarten, S.~M., et al.\ 2009, \apj, 691, 115 


\bibitem[{{Goudfrooij} {et~al.}(1994)}]{gou+94}
{Goudfrooij}, P. {et~al.} 1994, \aaps, 105, 341

\bibitem[Hakobyan et al.(2008)]{hak+08} 
Hakobyan, A.~A. {et~al.} 2008, \aap, 488, 523 

\bibitem[{{Ho} {et~al.}(1997){Ho}, {Filippenko}, \& {Sargent}}]{ho+97a}
{Ho}, L.~C., {Filippenko}, A.~V., \& {Sargent}, W.~L.~W. 1997, \apjs, 112, 315

\bibitem[{{Holtzman} {et~al.}(1995)}]{hol+95}
{Holtzman}, J.~A. {et~al.} 1995, \pasp, 107, 1065

\bibitem[{{Kawabata} {et~al.}(2010)}]{kaw+10}
{Kawabata}, K.~S. {et~al.} 2010, \nat, 465, 326

\bibitem[{{Kennicutt}(1998)}]{ken98}
{Kennicutt}, Jr., R.~C. 1998, ARA\&A, 36, 189

\bibitem[{{Kobulnicky} \& {Fryer}(2007)}]{kob+07}
{Kobulnicky}, H.~A. \& {Fryer}, C.~L. 2007, ApJ, 670, 747

\bibitem[{{Kroupa}(2001)}]{kro01}
{Kroupa}, P. 2001, MNRAS, 322, 231

\bibitem[{{Kundu} {et~al.}(1999)}]{kun+99}
{Kundu}, A. {et~al.} 1999, \apj, 513, 733

\bibitem[{{Le Borgne} \& {Rocca-Volmerange}(2002)}]{leb+02}
{Le Borgne}, D. \& {Rocca-Volmerange}, B. 2002, \aap, 386, 446

\bibitem[{{Leonard} \& {Duncan}(1990)}]{leo+90}
{Leonard}, P.~J.~T. \& {Duncan}, M.~J. 1990, AJ, 99, 608

\bibitem[{{Maund} \& {Smartt}(2005)}]{mau+05}
{Maund}, J.~R. \& {Smartt}, S.~J. 2005, MNRAS, 360, 288

\bibitem[{{Meynet} \& {Maeder}(2005)}]{mey+05}
{Meynet}, G. \& {Maeder}, A. 2005, \aap, 429, 581

\bibitem[{{Moellenhoff} \& {Bender}(1989)}]{moe+89}
{Moellenhoff}, C. \& {Bender}, R. 1989, \aap, 214, 61

\bibitem[{{Neill} {et~al.}(2009)}]{nei+09}
{Neill}, J.~D. {et~al.} 2009, \apj, 707, 1449

\bibitem[{{Perets}(2009)}]{per09b}
{Perets}, H.~B. 2009, ApJ, 698, 1330

\bibitem[{{Perets} \& Subr(2010)}]{per+10b}
{Perets}, H.~B. \& Subr, L. 2010, In preparation

\bibitem[{{Perets} {et~al.}(2010)}]{per+10}
{Perets}, H.~B. {et~al.} 2010, \nat, 465, 322

\bibitem[{{Pflamm-Altenburg} {et~al.}(2007){Pflamm-Altenburg}, {Weidner}, \&
  {Kroupa}}]{pfl+07}
{Pflamm-Altenburg}, J., {Weidner}, C., \& {Kroupa}, P. 2007, \apj, 671, 1550

\bibitem[{{Rich} {et~al.}(2005)}]{ric+05}
{Rich}, R.~M. {et~al.} 2005, \apjl, 619, L107

\bibitem[{{Sandage} \& {Visvanathan}(1978)}]{san+78}
{Sandage}, A. \& {Visvanathan}, N. 1978, \apj, 225, 742

\bibitem[{{Schilbach} \& {R{\"o}ser}(2008)}]{sch+08}
{Schilbach}, E. \& {R{\"o}ser}, S. 2008, A\&A, 489, 105

\bibitem[Schaller et al.(1992)]{sch+92} 
Schaller, G., Schaerer, D., Meynet, G., \& Maeder, A.\ 1992, \aaps, 96, 269 

\bibitem[{{Schlegel} {et~al.}(1998){Schlegel}, {Finkbeiner}, \&
  {Davis}}]{sch+98}
{Schlegel}, D.~J., {Finkbeiner}, D.~P., \& {Davis}, M. 1998, \apj, 500, 525

\bibitem[Schweizer et al.(1990)]{sch+90} 
Schweizer, F. {et~al.} 1990, \apjl, 364, L33 

\bibitem[{{Shen} {et~al.}(2010)}]{she+10}
{Shen}, K.~J. {et~al.} 2010, \apj, 715, 767

\bibitem[{{Sirianni} {et~al.}(2005)}]{sir+05}
{Sirianni}, M. {et~al.} 2005, \pasp, 117, 1049

\bibitem[{{Skrutskie} {et~al.}(2006)}]{skr+06}
{Skrutskie}, M.~F. {et~al.} 2006, \aj, 131, 1163

\bibitem[{{Smith} \& {Maeder}(1989)}]{smi+89}
{Smith}, L.~F. \& {Maeder}, A. 1989, \aap, 211, 71

\bibitem[{{Sullivan} {et~al.}(2006)}]{sul+06}
{Sullivan}, M. {et~al.} 2006, \apj, 648, 868

\bibitem[{{Sullivan} {et~al.}(2010)}]{sul+10}
---. 2010, \mnras, 755

\bibitem[{{Waldman} {et~al.}(2010)}]{wal+10}
{Waldman}, R. {et~al.} 2010, ArXiv:1009.3829

\bibitem[{{Wyder} {et~al.}(2007)}]{wyd+07}
{Wyder}, T.~K. {et~al.} 2007, \apjs, 173, 293

\bibitem[{{Yi} {et~al.}(2005)}]{yi+05}
{Yi}, S.~K. {et~al.} 2005, \apjl, 619, L111

\end{thebibliography}

\end{document}